\documentclass[aps, twocolumn, nofootinbib]{revtex4-1}

\usepackage{graphicx}
\usepackage{amsmath}
\usepackage{dsfont}
\usepackage{amssymb}
\usepackage{physics}
\usepackage{hyperref}
\usepackage{ulem}
\hypersetup{colorlinks=true,
	    final=true,
	    linkcolor=blue,
	    citecolor=blue,
	    filecolor=blue,
	    urlcolor=blue,}
\usepackage{amssymb}
\usepackage{subfigure}

\newcommand{\pr}{\partial}
\newcommand{\rta}{\rightarrow}

\newcommand{\ep}{\epsilon}

\newcommand{\p}{\prime}

\newcommand{\ra}{\rangle}

\newcommand{\beq}{\begin{equation}}
\newcommand{\eeq}{\end{equation}}

\newcommand{\upa}{\uparrow}

\newcommand{\downa}{\downarrow}

\newcommand{\barray}{\begin{eqnarray}}
\newcommand{\earray}{\end{eqnarray}}
\newcommand{\bwt}{\begin{widetext}}
\newcommand{\ewt}{\end{widetext}}

\begin{document}

\title{40 years of cuprate high-Tc  superconductors: a perspective on theories}
\author{Navinder Singh Bathinda}
\email{navinder.phy@gmail.com; Phone: +91 9662680605}
\affiliation{Theoretical Physics Division, Physical Research Laboratory (A Unit of the Department of Space, Government of India), Navrangpura Campus,  Ahmedabad, India. PIN: 380009.}

\begin{abstract}
An attempt is made to give a brief but coherent account of the situation of the theoretical ideas in addressing the mechanism of superconductivity in cuprate high-Tc superconductors. Specifically, the idea of superconductivity from repulsive interactions is discussed as it is gaining ground since the {\it consensus} paper was written in 2015\cite{kei}. The challenges it faces are also discussed. Three main schools of thought are presented, and an experimental result of 2022 pertaining to Anderson's super-exchange mechanism is also discussed.  An updated list of Anderson's ``dogmas" is also presented, as after year 2000, many other universally applicable experimental facts have been discovered. The ``dogmas" are universal facts which are distilled from a variety of complex experimental results, and highlights the key findings that seem to be central to the mechanism of superconductivity in cuprates. These are discussed as a commemoration of 40 years of high-Tc cuprate research. 
\end{abstract}

\maketitle

%
%
%
%

\section{Introduction}

The cuprate high-Tc superconductors are turning 40 years old this year! In this perspective article, I would like to present a brief survey of the theoretical landscape. Five years ago a critique article was written by the author\cite{nav1}. The current article is an extension of that article and attempts to present an updated account of the field, specifically, it adds: (1) a perspective on the three main schools of thought, (2) the problems that those face, and (3) an updated list of Anderson's ``dogmas" (section V) along with a brief discussion of an important experimental result of 2022 (section II C). Given the contentious nature of the field, every effort is made to maintain strict objectivity by grounding all arguments in well-established experimental facts. As the literature is vast, the author apologizes at the outset if any key papers are inadvertently omitted from the citations.

P. W. Anderson\cite{ander1}, while arguing against the pairing glue motivated ideas for the solution of the problem of high-$T_c$ superconductivity in cuprates, writes that the need for a bosonic glue to pair electrons is folklore! He alludes to the two basic mechanisms of pairing:

\begin{enumerate}

\item The Pitaevskii--Brueckner mechanism (known as superconductivity from repulsion in the contemporary literature).

\item The dynamical screening by a slow bosonic mode (like phonons in the case of BCS superconductors). 

\end{enumerate}

The first idea from the list above, which is connected with superconductivity from repulsion and pairing in higher momentum channels, has gained significant ground since the 'consensus' article was written in 2015\cite{kei}.  The question is whether {\it superconductivity from repulsion} is a good framework to understand unconventional superconductivity? Whether it is able to address the key experimental features?

Before we take up this topic,  let us first have a look at the history and see how this idea emerged.  It turns out that it was Lev Landau who proposed this idea (figure (1)). L. P. Pitaevskii  in his fundamental paper\cite{pita} entitled ``On the superfluidity of liquid $^3$He" writes:

\begin{quote}
{\texttt{``As Landau has pointed out, for the formation of Cooper pairs in a Fermi system it is sufficient that the interaction between the elementary excitations have an attractive character for any one value of the relative angular momentum $l$.....in this case the pairs will be formed with non-zero orbital momentum." } }
\end{quote}

Thus Landau pointed out that superconductivity can happen if any one component (with a specific relative angular momentum component $l$) of the interaction is negative (attractive), and the pairs will form with non-zero value of the relative angular momentum, {\it even though the total interaction can be repulsive}. These investigations were in the context of liquid $^3He$. In $^3He$, atoms are fermions and superfluidity happens in it at $\sim 2~mK$ in the p-wave channel. 

\begin{figure}[htbp]
  \centering
    \includegraphics[width=1.0\columnwidth]{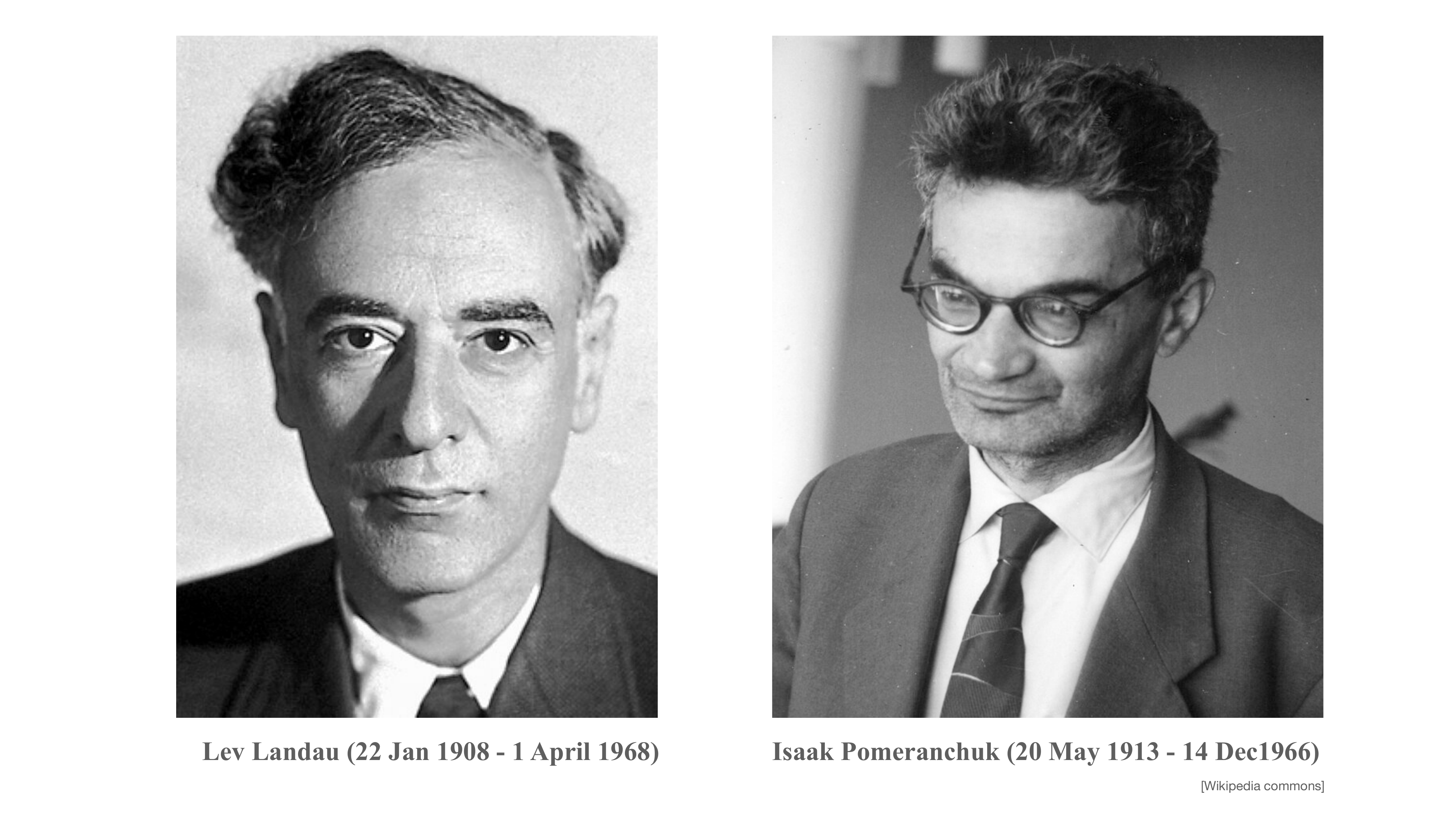}
    \caption{Lev Landau and Isaak Pomeranchuk.}
  \label{fig1}
\end{figure}

Question is, what causes pairing in $^3He$? The simplest way to understand the basic principle under action is through Pitaevskii--Brueckner mechanism (to which P. W. Anderson referred to in the context of Cuprates in his article\cite{ander1}). The interaction between two helium atoms can be schematically represented as depicted in figure (2).  At a shorter distance (compared to the mean distance $r_0$) helium atoms repel each other, but when the distance is greater than the mean distance ($r_0$), a weak attraction emerges. Now, the higher angular momentum channel (p-wave in the case of $^3He$) leads to a larger distance between the helium atoms where weak attraction emerges which further leads to pairing of the helium atoms into Cooper pairs in that channel. This aspect is beautifully expressed by Pomeranchuk (figure 1). Pitaevskii writes\cite{pita}: 

\begin{quote}
{\texttt{``.....Pomeranchuk has remarked that at sufficiently large values of the angular momentum, the excitation should evidently attract one another. This is because the large values of the momentum correspond to large distances between the excitations, so that forces should act which are analogous to the van der Waals forces between distant atoms." } }
\end{quote}

\begin{figure}[htbp]
  \centering
    \includegraphics[width=1.0\columnwidth]{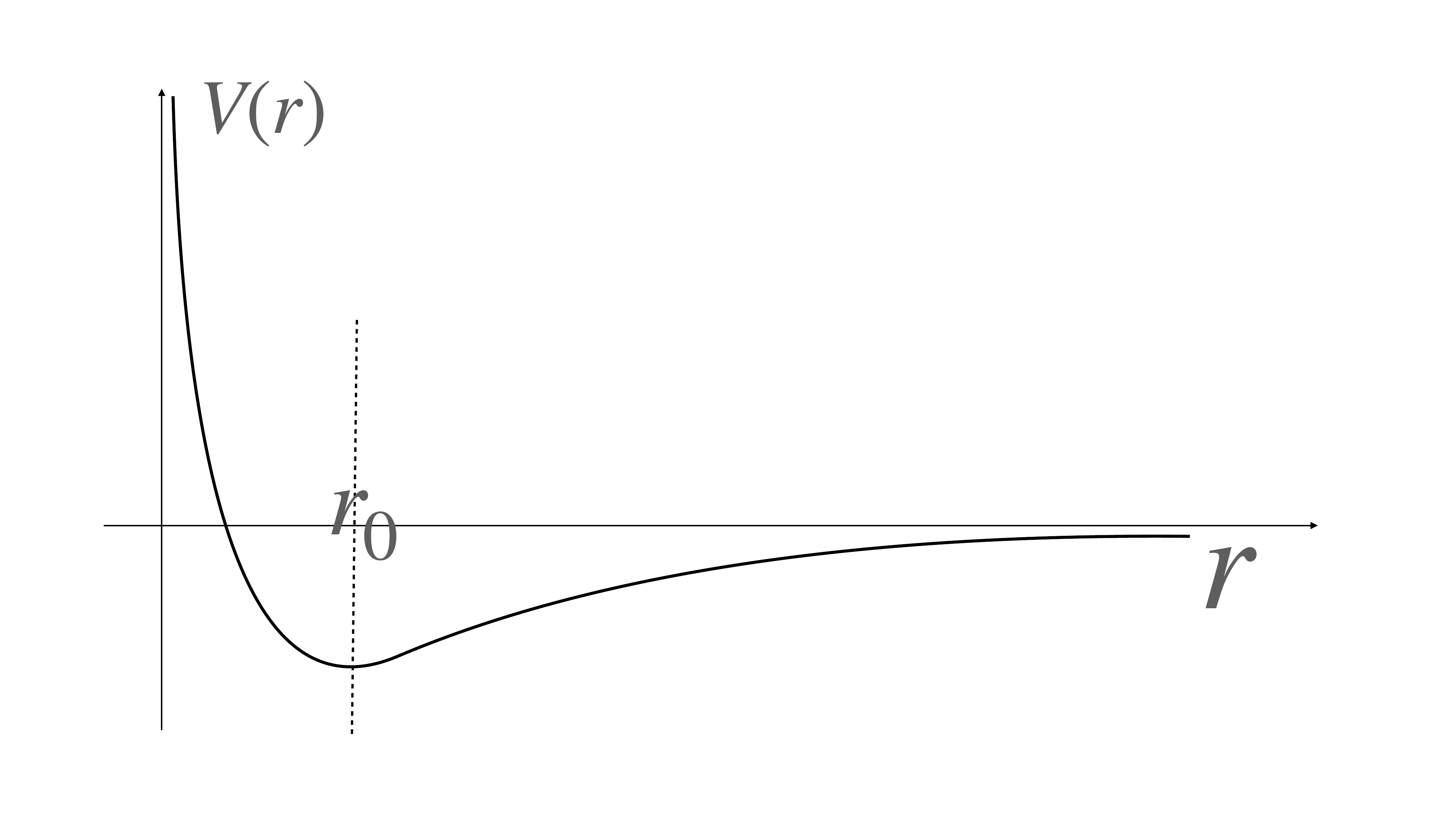}
    \caption{Schematic interaction between two helium atoms.}
  \label{fig2}
\end{figure}

In other words, higher angular momentum channels lead to larger average separation which in turn induces attractive interaction. Pomeranchuk's argument should be viewed as providing a physical mechanism of attraction in the scheme proposed by Landau (the pairing problem de-couples in the angular momentum channels, and it is sufficient to have attraction in any one of the components).\footnote{It turns out that $^3He$ case is more complex and interesting as its atoms have a finite magnetic moment.  When external magnetic field is applied, the  spin configurations $|\upa \upa\ra$ and $|\downa\downa\ra$ are selected out (the $A'-phase$ depending on the direction of the applied magnetic field) whereas in the B-phase all the possible spin configurations exist ($|\upa \upa\ra,~|\downa\downa\ra,~~\frac{1}{\sqrt{2}}(|\upa\downa\ra + |\downa\upa)$) with equal weight. In the A-phase because of the ferromagnetic fluctuations at higher pressure  $|\upa\upa\ra$ and $|\downa\downa\ra$ have more weight than that of $\frac{1}{\sqrt{2}}(|\upa\downa\ra + |\downa\upa\ra)$ thus the former are selected out. The scenarios for the superfluidity of $^3He$ are: For A-phase: the Anderson--Brinkman--Morel; and for the B-phase:  the Balian--Werthamer, and in addition paramagnonic contributions also play the role. Refer to\cite{tony, voll, leggett1} for more details.}

The idea of pairing from purely repulsive electronic interactions was further taken up by  W. Kohn and J. M. Luttinger in 1965\cite{kohn}. In fact, an analog of Pomeranchuk's argument exists in metallic systems. The argument here is that the Friedel oscillations can provide an attractive part in otherwise repulsive interaction. In real space, the attractive component of the screened Coulomb repulsion emerges at large distances.  Translating it into momentum space, it turns out that the interaction has attractive components for large angular momentum quantum numbers.  More precisely, in isotropic systems, the fully screened Coulomb repulsion can have attractive components for large odd values of the angular quantum numbers\cite{kohn}.

However, we must note two central points:

\begin{enumerate}

\item {\it The spin-fluctuations must be seen as a ``selector" or ``prioritizer" of the spin configuration of the Cooper pairs, not  only as a slow boson mode providing the pairing glue in the sense of the BCS theory.}

\item  Is it necessary for superconductivity to happen that a physical mechanism of attractive interaction be present? Such as the requirement of the Pomeranchuk argument in the case of $^3\text{He}$ that, at larger interatomic distances (higher angular momentum channels), a weak attraction emerges. And for electrons in systems having a sharp Fermi surface, the Kohn-Luttinger argument that Friedel oscillations can provide an attractive component within an otherwise purely repulsive interaction for large angular momentum channels? 

\end{enumerate}

The first point is an old idea by Fay and Layzer\cite{lay}, but has become very relevant in the recent advances of unconventional superconductivity (In Nickelates, Iron based superconductors, Cuprates etc).

As to the second point it turns out that physical mechanism of attraction at higher angular momentum is not always required.  Regarding the state of electrons in cuprates it is not possible to apply Pomeranchuk's argument (interaction between two electrons is not van der Waals interaction, and at larger values of relative angular momentum it does not turn into a weak attraction). The interaction is manifestly repulsive (the screened Coulomb repulsion). In addition, magnetic fluctuations are omnipresent in the entire doping range\cite{kei}. Application of the Fermi-liquid theory, and thus Kohn--Luttinger ideas (Friedel oscillations originating from sharp Fermi surface) is not beyond criticism when applied to cuprates especially in the underdoped side of the superconducting dome\cite{nav1}. {\it  Strong correlations and magnetic fluctuations so dominantly  govern the state of electrons in cuprates. } For more physical discussion for the mechanism of pairing via repulsive interactions refer, for example, to\cite{kei}.

The rest of the paper is organized in the following way. Before we discuss the hurdle that mechanism of superconductivity from repulsion faces, we would like to review three main camps which attempt to address the problem of pairing in cuprates in section II. Then at the end of this section we will collect evidence which support the notion of an overarching theory. Then the question is asked: Whether the mechanism of superconductivity from repulsion is a good candidate for an overarching theory? In section II we also discuss the experimental verification of c-DMFT result which is connected with Anderson's super-exchange mechanism. In section III we discuss the hurdle that these proposed mechanisms face. Then in section IV, the original list of Anderson's ``dogmas" is given, and in section V an updated list is presented as after year 2000, many more universally applicable experimental facts have been discovered. We conclude the draft by presenting a list of the summary points in section VI.

\section{The three main schools of thought}

\subsection{The pairing glue--soft modes?}


It is argued that as cuprates and other unconventional superconductors are near magnetic instabilities (as magnetism weaken with doping, pressure, etc.,  superconductivity in the form of a dome appears), the magnetic spin fluctuations of the lost/weakened magnetic phase can provide the pairing glue similar to the case of the BCS theory (where phonon mediated retarded interaction provides the required glue).  It is argued that the magnetic spin fluctuations (with more complex frequency and wave-vector dependence) can do the job in the case of cuprates and other unconventional superconductors)\cite{scala}.

For the single band Hubbard model a conceptual difficulty is generally raised\cite{kei,nav1}. {\it How can the same electrons that are being paired with each other also provide the very pairing glue?}. But this conceptual difficulty can be easily cleared (apparently!): long wavelength collective fluctuations of electronic degrees-of-freedom itself can act as the required bosons that can execute transitions of Cooper pairs between different momentum levels much like in the case of BCS superconductors where virtual phonons execute these transitions. 

But this line of reasoning runs into a different problem as pointed out by Anthony Leggett in his Nobel lecture\cite{leggett1}. As more and more electrons bind into Cooper pairs and go into the condensate, the very mechanism of the generation of long wavelength collective fluctuations in the normal electron gas itself weakens (as the normal component reduces and the superfluid/superconducting component increases as the temperature is lowered below the critical temperature). The question is how do the glue itself survives the condensation process?   Whereas in the case of BCS superconductors, phonon degrees-of-freedom (which is a different sub-system) are hardly affected by the condensation process!

In addition, pertaining to the single band Hubbard model, P. W. Anderson raised another serious criticism\cite{ander1}: 

If the single-band Hubbard model with two parameters $t$ and $U$ (or its strong coupling version $t$ and $J \sim\frac{t^2}{U}$ model) is used to model cuprates, then Anderson argues that the exchange of AFM spin fluctuations (used as a pairing glue) is actually a very high-energy (high-frequency) phenomenon. Whereas the Migdal-Eliashberg extension of the BCS theory describes electron pairs bound together by an exchange of low-frequency bosons.  Both the scales $J$ and $U$ constitute a very high energy (high frequency) dynamics.  Anderson writes\cite{ander1}:

\begin{quote}
{\texttt{
``The crucial point is that there are two very strong interactions, $U ~(>2 ~eV)$ and $J(~0.12 eV)$, that we know are present in the cuprates, both a priori and because of incontrovertible experimental evidence. Neither is properly described by a bosonic glue, and between the two it is easy to account for the existence of antiferromagnetism, d-wave superconductivity, and many other phenomena of high-Tc superconductivity. Whether any additional ``glue" exists is of lesser interest." 
}}
\end{quote}

These criticisms are applied to the single band Hubbard model.  However, one can argue that the actual physics of the cuprates is much more complex (some aspects are beyond the scope of the single band Hubbard model). These questions remained unanswered to date. Recently, this author has carefully analyzed (with respect to experimental evidences) both the single-band and three-band models, and came to the conclusion that the three band (Emery) model is more appropriate (as far as the phenomenology of Barzykin-Pines and Gor'kov-Teitle'baum is concerned)\cite{nav2}. There is a vast literature on these two models, and referencing them all here is not possible (interested readers can browse the detailed references cited in\cite{nav2}). Thus, it seems that a more appropriate model is the three band Hubbard model (the Emery model), and Anderson's criticism may be addressed. But these comments remain debatable. In addition, the soft modes originating from the QCP has been postulated to be constituting the required slow modes (bosons) for pairing, along with the non-Fermi liquid behaviour\cite{fer1,fer2,fer3,fer4,fer5,fer6}.

\subsection{The purely repulsive interactions?}

In this scenario, Anderson's objection does not apply. With all repulsive interactions, unconventional superconductivity is argued to be possible through a non-trivial mechanism\cite{kei,raghu}. In the strong coupling limit of the Hubbard mode (Hubbard $U$ of the order of the bandwidth $W$) no analytic theory and no intuitive picture are available.  However, it can be intuitively motivated in the special case of weak coupling limit of the Hubbard model (where $U$ is much less than the bandwidth ($U<<W =8t$)). Though this limit is not directly applicable to the case of cuprates, but it can give us an idea of the basic mechanism involved.  The gap structure is inferred from the solution of the BCS equation:

\beq
\Delta_k = -\sum_{k^\p} \Gamma_{k,k^\p} \frac{\Delta_{k^\p}}{2 \sqrt{\ep_{k^\p}^2 +\Delta_{k^\p}^2}}
\eeq

Here $\Gamma_{k,k^\p}$ is an effective repulsive interaction (an appropriately renormalized two-particle vertex function)\cite{raghu,kei}. It is argued that if $\Gamma$ is sufficiently $k$ dependent function and peaks at a special wave vector $Q$, then a sign changing gap ($\Delta_{k+Q} = -\Delta_k$) is a solution of the above gap equation with pair-forming interactions that involve large momentum transfer near $Q$, whereas interactions in which momentum transfer is much smaller than $Q$ are in fact pair breaking. This again leads to the correct d-wave gap function (with $\Delta(k)$ and $\Delta(k+Q)$ have opposite signs). 

However, as noted\cite{kei} this nice conceptual picture is available only in the weak coupling limit. In the intermediate coupling regime ($U\sim t$), which is relevant for cuprates, no such conceptually transparent picture is available.

\subsection{Super-exchange, the super-glue?}

The simplest way to appreciate this idea is to understand Anderson's words\cite{ander1}: 

\begin{quote}
{\texttt{
``A second consequence of $U$ is the appearance of a large antiferromagnetic exchange
coupling $J$, which attracts electrons of opposite spins to be on neighboring sites.
.....Because of the large magnitude of $J$, the
pairing can be very strong, but only a fraction
of this pairing energy shows up as a superconducting $T_c$, for various rather complicated but
well-understood reasons."
}}
\end{quote}

Thus, according to Anderson, it is the superexchange interaction $J$ which is responsible for pairing. Another way to appreciate pairing of two holes due to the presence of short ranged AFM correlations is to consider the pairon model introduced by Yves Noat etal\cite{noat1,noat2,noat3}. 

While the hole doping destroys magnetic order, the key proposal for electron pairing is that it preserves pair-forming superexchange interactions, which are governed by the charge-transfer energy scale. 

There is an important advancement on the experimental side pertaining to this idea of pairing via super-exchange. The pairing scale is governed by $J$ which scales as $\frac{t^4}{\ep^3}$ as deduced from the three-band model. Here, $\ep$ is the charge transfer energy. Thus, reduced charge transfer energy should lead to stronger pairing (as $J\sim \frac{1}{\ep^3}$).

In a 2022 study\cite{mahony}, Shane M. O'Mahony and collaborators investigated this connection, and achieved a coterminous visualization of the local charge transfer energy $\ep(r)$ and local electron pair density $n_P(\ep)$ by measuring the dependence of these quantities on the distance of the apical oxygen atom from the planar copper atom (the height  $\delta(r)$ of the apical oxygen atom from the copper atom just below it). To get it, the authors used a classic high-temperature superconductor $\text{Bi}_2\text{Sr}_2\text{CaCu}_2\text{O}_{8+x}$ at optimal doping, $p = 0.17$. This material has a natural, wavy variation in its crystal structure (called a supermodulation, $\delta(r)$) that repeats every $26\text{ \AA}$ (that is, the height of apical oxygen has this periodicity). It turns out that the apical oxygen height variation modulates the charge transfer energy. 

The authors measured the differential conductance $g(r,V)$ as a function of the location $r$ and the tip-sample voltage $V$. From which they derived the spatially resolved EDOS.  The local charge transfer energy $\ep(r)$ is then derived as the minimum distance between upper and lower bands at constant conductance(refer to\cite{mahony} for details). Basically, a $\delta(r)$ dependence of $\ep(r)$ is determined. Next, using the superconducting scanning tunneling microscope (sup-STM) tips (of the same material), Josephson critical current (for electron pair tunneling) is also measured as a function of $r$ (and thus $\delta(r)$) from which electron pair density $n_P(r)$ is determined as a function of $\delta(r)$.  By eliminating the common variable $\delta(r)$ they find that the electron pair density decreases linearly as the charge transfer energy increases ($\frac{dn_P}{d\epsilon} \approx -0.81 \pm 0.17\text{ eV}^{-1}$). This value agrees reasonably well with results from c-DMFT theory\cite{mahony}.

However, the connection of the c-DMFT result with superexchange mechanism is very subtle\cite{weber,t9}\footnote{The cellular or cluster DMFT comprehensively captures short-range correlations within the designated cluster. Crucially, calculating the superconducting state does not rely on a conventional gap equation; instead, the anomalous self-energy is initially taken to be finite. If this finiteness persists upon self-consistently solving the effective impurity problem, it induces an anomalous component within the dynamical mean field, yielding a self-consistent superconducting state. Owing to this strict self-consistency requirement, this framework represents a far more rigorous methodology than approaches dependent on an a-priori superconducting ansatz\cite{nav1}.}. Referring to\cite{weber}, C. Weber and collaborators write:

\begin{quote}
{\texttt{
``For $\ep_d-\ep_p$ [the charge transfer energy], its large value in the strong correlation limit suppresses
charge-fluctuations, rendering the residual superexchange interaction between the doped holes weak, resulting in low
superconducting temperatures. As we decrease  $\ep_d-\ep_p$, superconducting tendencies increase as we pass through
the intermediate correlation regime, until we reach the weak correlation limit." 
}}
\end{quote}

It turns out that the height of the apex oxygen ion modulates the Charge Transfer Energy (CTE ), $\ep = \ep_d -\ep_p$,  in such a way that the increased height of the apex oxygen ion reduces $\epsilon_d$ thereby making CTE smaller in magnitude. This yields an enhanced superexchange interaction $J \sim \frac{1}{\ep^3}$, and thereby facilitates pairing.  Now, the height of the apex oxygen ion affects both $\ep_p$ and $\ep_d$. But its effect is more dominating in $\ep_d$ due to the change in the overlap of copper $3d_3z^2-r^2$ orbital with the apex oxygen orbital (hybridization effect). There is also the crystal field and electrostatic effects (refer to\cite{mahony} and citations therein).

This experimental verification is an important step regarding the link between super-exchange $J$ and $T_c$\cite{mahony}. However,  we need to see whether the community accepts it as the solution of the key problem of cuprates?  As Anderson put it, the superexchange $J$, which attracts electrons of opposite spins to neighboring sites, serves as the primary key to pairing. {\it How do we decide whether superconductivity from repulsion is the key or super-exchange is the key? To get insight into this question, it is useful to look at the problem from a broader perspective.}

\subsection{Arguments in favor of an overarching theory}

We postpone the questions raised above and follow another argument in the favor of an overarching theory (a common theme for cuprates (CupSCs), heavy fermion superconductors (HFSCs), iron-based superconductors (IBSCs), organic superconductors (OSCs) etc)\cite{scala, ste}. Basically, the following points were noted\cite{scala,ste}:

\begin{enumerate}

\item All these unconventional superconductors (CupSCs, IBSCs, HFSCs, OSCs, etc) are near to some sort of magnetic phase. When magnetism is weakened by doping/pressure etc, it gives a way to unconventional superconductivity (USC).

\item The pseudogap in cuprates has the same ``place" or ``role" as that of a magnetic state (AFM, SDW etc) in HFSCs, IBSCs etc.

\item Pairing is in the higher momentum states:

\item[] $S\pm$ for IBSCs.

\item[] $d-wave$ for CupScs.

\item[] $d-wave$ for OSCs.

\item[] $d-wave$ in many HFSCs.

\item[] $p-wave$ in $UGe_2$, $UBe_{13}$.

\item[] $f-wave$ in $UPt_3$ etc. 

\item DC resistivity above the dome is not phonon mediated. 

\end{enumerate}

The arguments in the favor of an overarching theory are quite strong, however, system-to-system variations would matter. It is difficult to understand that Anderson's super-exchange idea can form a common thread.  Or, the mechanism of superconductivity in these systems should be understood case by case? Can SC from repulsion be a candidate for an overarching theory? The similarities are so striking that we cannot neglect them.

\section{The biggest hurdle}

We set aside the question of the pairing mechanism, and consider an important experimental feature applicable on the overdoped side of the superconducting dome.  The following issue remains unaddressed, regardless of whether superconductivity is assumed to arise from the repulsive mechanism or from the superexchange mechanism.

There is a very central and crucial experimental observation that is directly linked to the very mechanism of pairing. This observation was made by Taillefer and collaborators\cite{taillefer1}. It turns out that from the overdoped side, the onset temperature of superconductivity scales with the strength of the anomalous normal-state scattering that makes DC resistivity ($\rho$) linear in temperature.  If $p_c$ is the doping at which superconducting dome touches the doping axis in the overdoped side, that is, for $p\ge p_c$, $T_c=0$, then, when $p<p_c$, $T_c\ne 0$ and a $T-$linear component appears in resistivity: $\rho = A T +B T^2$. The coefficient $A$ is zero for $p>p_c$ and non-zero for $p<p_c$. This is true not only in the case of cuprates but it is true more generally (in heavy fermion superconductors, in iron based superconductors, and in organic superconductors as documented in\cite{taillefer1}). {\it It turns out that $T_c \propto A$, i.e., stronger are the pairing correlations, the larger the value of $A$ becomes.} This is the key experimental observation.  In addition, it turns out that the T-linear component covers the entire SC dome, once superconductivity is suppressed with sufficiently strong magnetic field (known in the contemporary literature as ``the elephant foot")\cite{cooper}.

Now, the mechanism of resistivity requires the presence of a bosonic momentum sink (in metals this bosonic momentum sink can be the phonon degrees-of-freedom). The above central observation says that this bosonic momentum sink is actually tied up (via scattering and pairing) with the very pairing mechanism (the larger $A$ is, the larger $T_c$ becomes). 

The question is: How do we reconcile the idea of superconductivity from repulsion or superconductivity from super-exchange interactions with this very central experimental observation (at least on the overdoped side)? The ideas based on soft modes--as a pairing glue -- suffer from some serious issues as discussed in section (II).  In this author's opinion, the experimentally derived relation $T_c\propto A$ would be the key test for any candidate theory. {\it The future challenge for the theory of unconventional superconductivity is precisely the rationalization of $T_c\propto A$.}




\renewcommand{\labelenumi}{$\blacklozenge~${\arabic{enumi}}}

\section{Anderson's ``dogmas" -- The original list}

We leave the main arguments of the paper, and as a commemoration of 40 years of high Tc cuprates, we re-visit Anderson's ``dogmas"\cite{anderb} and compile an updated and corrected list, as after year 2000 many more universally applicable experimental features have been discovered.

Derived from a thorough synthesis of pivotal experiments and tight logical reasoning, Anderson's ``dogmas" represent fundamental factual inferences \cite{anderb}. Their purpose is to strip away non-essential information and reduce the theoretical problem of the mechanism of superconductivity to its core components. These ``dogmas" are structured as follows:

\begin{enumerate}

\item
Anderson argued that all relevant spin and charge carriers reside within the antibonding $\text{Cu } 3d_{x^2-y^2} - \text{O } 2p_{\sigma}$ orbital network. Strong on-site Coulomb repulsion $U$ splits this band into upper and lower Hubbard bands, while all other electronic bands remain completely filled and thus electronically inert. Direct experimental validation for this first ``dogma" is provided by Batlogg's classic measurements -- as well as more recent observations\cite{bari} -- of a universal in-plane resistivity ($\rho_{ab}$) per copper-oxide layer. This fundamentally establishes the two-dimensional nature of the cuprate problem \cite{anderb}

\item
Interplay of Magnetism and Superconductivity: Magnetism and superconductivity in the cuprates are intimately connected. The identical electrons responsible for antiferromagnetic (AFM) ordering at zero doping ($x=0$) transform into mobile charge carriers upon finite doping, ultimately condensing into Cooper pairs to exhibit superconductivity below a critical temperature $T_c$. In defense of this view, Anderson points out that the undoped parent compounds are classic Mott-Hubbard insulators exhibiting long-range AFM order. This microscopic picture is further corroborated by a wealth of compelling data from optical conductivity, photoemission spectroscopy, and nuclear magnetic resonance (NMR) measurements \cite{anderb}

\item
Repulsive Interactions as the Dominant Scale: The dominant microscopic interactions within the copper-oxide planes are repulsive, characterized by large energy scales where $U$ and $J$ dictate the core physics. Consequently, alternative mechanisms -- such as electron-phonon couplings or electron-impurity interactions--play a sub-dominant role. This hierarchy is fundamentally rooted in the fact that, at zero doping, the cuprates form Mott-Hubbard insulators rather than standard band insulators\cite{anderb}.

\item
The Non-Fermi liquid behaviour: The electronic state (the ``normal" state from which superconductivity emerges) in copper-oxide planes is not a Fermi liquid in the sense that quasiparticle weight goes to zero ($Z\rta0$). In fact, by a tight comparison of the Marginal Fermi Liquid Theory (MFLT) with experiments\cite{verma89}, the following behaviour of the self-energy was inferred:
\beq
\Sigma^{''}\propto \omega,~~~~\Sigma^{'} \propto \omega \log(\omega),
\eeq
and  thus $Z = (1-\frac{\pr \Sigma}{\pr \omega})^{-1} \rta 0$. Therefore, the quasiparticles cannot be defined in the Landau sense.  Anderson advances several arguments to support it\cite{anderb}. Leading ones are the anisotropy in c-axis and ab-plane transport, and several conclusions from ARPES and optical measurements\cite{anderb}.

Anderson's "dogma" 5 and 6 are related to his {\it interlayer hopping} theory, which has turned out to be wrong, and criticized by him later\cite{anderph}.

\end{enumerate}


\section{Anderson's ``dogmas" -- An updated and extended list}

There is one important change one must consider in ``dogma" 1:  The oxygen orbitals cannot be integrated out. Hole doping creates holes in oxygen $p_{\sigma}$ orbitals (there is a direct evidence for it\cite{gau}). Within the scenario of one-band model many key experimental features (such as Johnston-Nakano scaling etc) cannot be rationalized. The Gor'kov - Teitle'baum and Barzykin-Pines phenomenology find clear explanation in the two band Emery model (refer to\cite{nav2} for details and refer to citations therein).

Next, we collect other universally applicable (not restricted to one Cuprate member) experimental facts, and extend the list of essential ``dogmas". In this author's opinion these also constitute the ``core" of the problem and related issues. 

\begin{enumerate} 

\setcounter{enumi}{4}

\item  It turns out that -- even though the finite temperature normal state (strange metal phase) exhibits anomalous (non-Fermi liquid) properties --  the ground state of Cuprates has been found to be A Fermi liquid\cite{taillefer2}. The obedience  of the Wiedemann-Franz law and the obedience of Lifshitz--Kosevich formula show that the ground state ($T\rta0$) of cuprates is a Fermi liquid\cite{taillefer2}. No exotic physics is needed! It clearly shows that the strange metal phase emerges out of a Fermi liquid due to anomalous scattering (finite temperature effects of the strong dominance of magnetic fluctuations and strong electronic correlations). This is another very hard problem: How does finite temperature non-Fermi liquid behaviour emerge from a Fermi liquid ground state?

\item
As discussed in the ``biggest hurdle" section, Taillefer and collaborators showed that, from the overdoped side, the onset temperature of superconductivity scales with the strength of the anomalous normal-state scattering that makes DC resistivity ($\rho$) linear in temperature\cite{taillefer1}. 

Within the dome regime ($p < p_c$), a finite transition temperature $T_c \neq 0$ emerges, accompanied by the onset of a $T$-linear component in the low-temperature resistivity: $\rho = A T + B T^2$. The coefficient $A$ vanishes identically for $p > p_c$ and remains finite for $p < p_c$. 
This empirical trend is not unique to the cuprates; rather, it represents a universal hallmark across diverse unconventional superconductors, including heavy-fermion, iron-based, and organic superconductors, as comprehensively compiled by Taillefer \cite{taillefer1}.  Crucially, this manifests a distinct scaling relation: $T_c \propto A$ which implies that the strength of the pairing correlations directly scales with the magnitude of the linear-in-$T$ scattering coefficient.

\item Interesting physics around $p^*\simeq 0.19$ (note that $p_c$ and $p^*$ are different points)?

As $T \rightarrow 0$, there is a PseudoGap Quantum Critical Point (PG QCP) which is located at $p^*$\cite{taillefer2}. Below the critical doping ($p < p^*$), DC resistivity shows a pronounced upturn at low temperatures before eventually saturating. As emphasized by Louis Taillefer, this behavior marks a clear metal-to-metal transition. This low-temperature rise in resistivity stems from a sharp loss of charge carriers as the system enters the pseudogap phase. Hall coefficient measurements directly support this picture\cite{taillefer2}. Specifically, the carrier density drops from $1+p$ holes per $\text{Cu}$ atom when $p > p^*$ down to just $p$ holes per $\text{Cu}$ atom when $p < p^*$. It leads to the central Mystery: Why does the carrier number drop so abruptly below $p^*$? The answer to this pivotal question holds the ultimate secret of the pseudogap phase (a somewhat artistic description is given in\cite{unhappy}). It could be due to quasi-localization of holes in the copper d-orbitals. The hints of this were already discussed in a very important but unpublished article by David Pines entitled ``Pseudogap behavior in underdoped cuprates", Cond-matt/0404151 (2004).

The scalings $C_{el} \propto -T\ln T$ and $\rho\propto T$ (logarithmic electronic specific heat and T-linear resistivity) are the typical signatures of AFM QCP\cite{taillefer2}.

These findings are corroborated by the behaviour of c-axis optical conductivity. It turns out that the Drude peak (c-axis) appears for $P \gtrsim 0.19$ which means that the coherence sets-in leading to ``traditional" metallic behaviour. The release of quasi-localized copper holes and enhancement of total carrier density to $1+p$ holes per copper atom is clearly seen in it\cite{tallon}.

In addition, there is another indirect evidence from induced magnetic moments on doped Zn impurities. These induced local moments can be seen to progressively fall to zero around $p\simeq 0.19$ again confirming Taillefer and collaborators observations' of $p$ to $1+p$ transition around $p^*$ as doping is increased through $p^*$. It also means that the PG is linked to short-ranged AFM correlations\cite{tallon}.

\item Phenomenology of the PG state (charge degrees-of-freedom): The Gor'kov-Teitel'baum thermal activation model of 2006 is a very useful phenomenology of the PG state of cuprates\cite{gtta1}. In 2006, Gor'kov and Teitel'baum developed a phenomenological model to describe the temperature and doping dependence of the Hall coefficient, $R_H(T)$, in $\text{La}_{2-x}\text{Sr}_x\text{CuO}_4$ \cite{gtta1}. They rationalized the experimental Hall data by introducing a temperature- and doping-dependent effective carrier density, $n_{\text{Hall}}(x,T)$, defined via the standard relation: $R_H(T) = \frac{1}{n_{\text{Hall}}(x,T)e}$. Within the Gor'kov-Teitel'baum framework, this effective carrier density is parameterized as:

\[n_{\text{Hall}}(x, T) = n_0(x) + n_1(x)\exp\left[-\frac{\Delta(x)}{T}\right]\]

where the total carrier density is decomposed into a temperature-independent component, $n_0(x)$, and a thermally activated, temperature-dependent component governed by $n_1(x)\exp[-\Delta(x)/T]$. This formulation establishes the governing equation of the Gor'kov-Teitel'baum Thermal Activation (GTTA) model, which successfully reproduces the experimental Hall effect data across a wide range of parameters in $\text{La}_{2-x}\text{Sr}_x\text{CuO}_4$ \cite{gtta1}. The evolution of the extracted parameters $n_0(x)$, $n_1(x)$, and $\Delta(x)$ yields key insights into the underlying physics: The Temperature-Independent Term, $n_0(x)$: For low doping levels ($x < 0.07$), $n_0(x)$ scales linearly with the nominal hole doping concentration (as if the doped holes are the only carriers). Beyond $x = 0.07$, however, it deviates significantly from this linear regime (more charge carriers join the carrier density).

The pre-exponential Factor, $n_1(x)$:  In the underdoped and optimally doped regimes ($x < 0.19$), $n_1(x)$ remains approximately constant ($\simeq 2.8$). It exhibits a sharp, abrupt decrease for $x > 0.19$, signaling a fundamental electronic phase reconstruction in the overdoped regime\cite{gtta1,gtta2}. This is connected with the interesting physics at $p^*$ as discussed above.

The thermal activation energy, $\Delta(x)$: The parameter $\Delta(x)$ represents the characteristic thermal activation energy of the charge carriers. Physically, it is interpreted as the effective energy gap between electronic states in antinodal regions of the Brillouin zone. For $x < 0.19$, $\Delta(x)$ decreases linearly with increasing doping, demonstrating exceptional quantitative agreement with the pseudogap scales directly extracted from angle-resolved photoemission spectroscopy (ARPES) measurements \cite{gtta1,gtta2}. It turns out that the GTTA model has much broader applicability and many PG related issues can be understood\cite{gtta2,gtta3}.  Therefore, as far as charge degrees-of-freedom are concerned GTTA model is a very successful phenomenology of the PG phase.

\item Phenomenology of the PG state (spin degrees-of-freedom):  The Barzykin-Pines phenomenological model.

The Barzykin-Pines two-component model is based on NMR Knight shift and static magnetic susceptibility ($\chi(x,T)$) data. Barzykin and Pines \cite{bp} proposed that underdoped cuprates comprise two distinct, interacting components: The Spin Liquid (SL) Component: (1) A 2D Heisenberg system of quasi-localized copper spins, and (2) mobile holes on the oxygen sub-lattice, which disrupt and weaken the nearest-neighbor copper superexchange coupling, leading to a doping-dependent effective exchange interaction: $J_{\text{eff}}(x) = J f(x), \quad \text{where}~ J \sim 1200\text{ K and } f(x) = 1 - \frac{x}{0.2}$.

This SL component exists below a crossover temperature $T_{\text{max}}$ -- the temperature where $\chi(x,T)$ reaches its maximum -- and vanishes abruptly above the quantum critical point ($p \simeq 0.2$).

The Fermi Liquid (FL) component is formed by the mobile holes residing in the oxygen $p$ orbitals, as described by Emery\cite{emery1, nav2}. This component grows with doping as $1 - f(x) = x/0.2$ and contributes a temperature-independent Pauli susceptibility. Consequently, the total magnetic susceptibility is a weighted superposition of both behaviors:

\[\chi(x,T) = f(x) \chi_{\text{SL}}(T) + (1-f(x))\chi_{\text{FL}}\]

This framework beautifully rationalizes Johnston-Nakano scaling(refer to\cite{bp,nav2}, and references therein). When susceptibility data across various doping levels $x$ are plotted as $\frac{\chi(T,x)}{\chi_{\text{max}}(x)}$ versus $\frac{T}{T_{\text{max}}(x)}$, they collapse onto a single, universal curve:  $\chi(T,x) = \chi_0(x) + [\chi_{\text{max}}(x) - \chi_0(x)] \mathbb{F}\left(\frac{T}{T_{\text{max}}(x)}\right)$.  The universal scaling function $\mathbb{F}$ matches the theoretical profile of a 2D Heisenberg model. Crucially, the temperature-dependent term originating from the localized SL component vanishes at $x \gtrsim 0.2$. This perfectly tracks the experimental $p$ to $1+p$ Fermi surface reconstruction at $p^* \simeq 0.19$\cite{taillefer2,sun}.

\item Main features of pseudogap around $p^*$: (1) An anti-nodal (along Cu-O direction) spectral gap opens below $p^*$; (2) EDOS decreases below $p^*$; (3) Charge carrier density drops below $p^*$ ($1+p$ to $p$ transition); (4) at $p^*$ the signatures $C_{el} \propto-T ln T, ~~\rho\propto T$ are that of AFM quantum criticality, thus PG is connected with short ranged AFM correlations; (5) the ground state of the PG is a Fermi liquid\cite{taillefer2}. 

\begin{figure}[htbp]
  \centering
    \includegraphics[width=1.0\columnwidth]{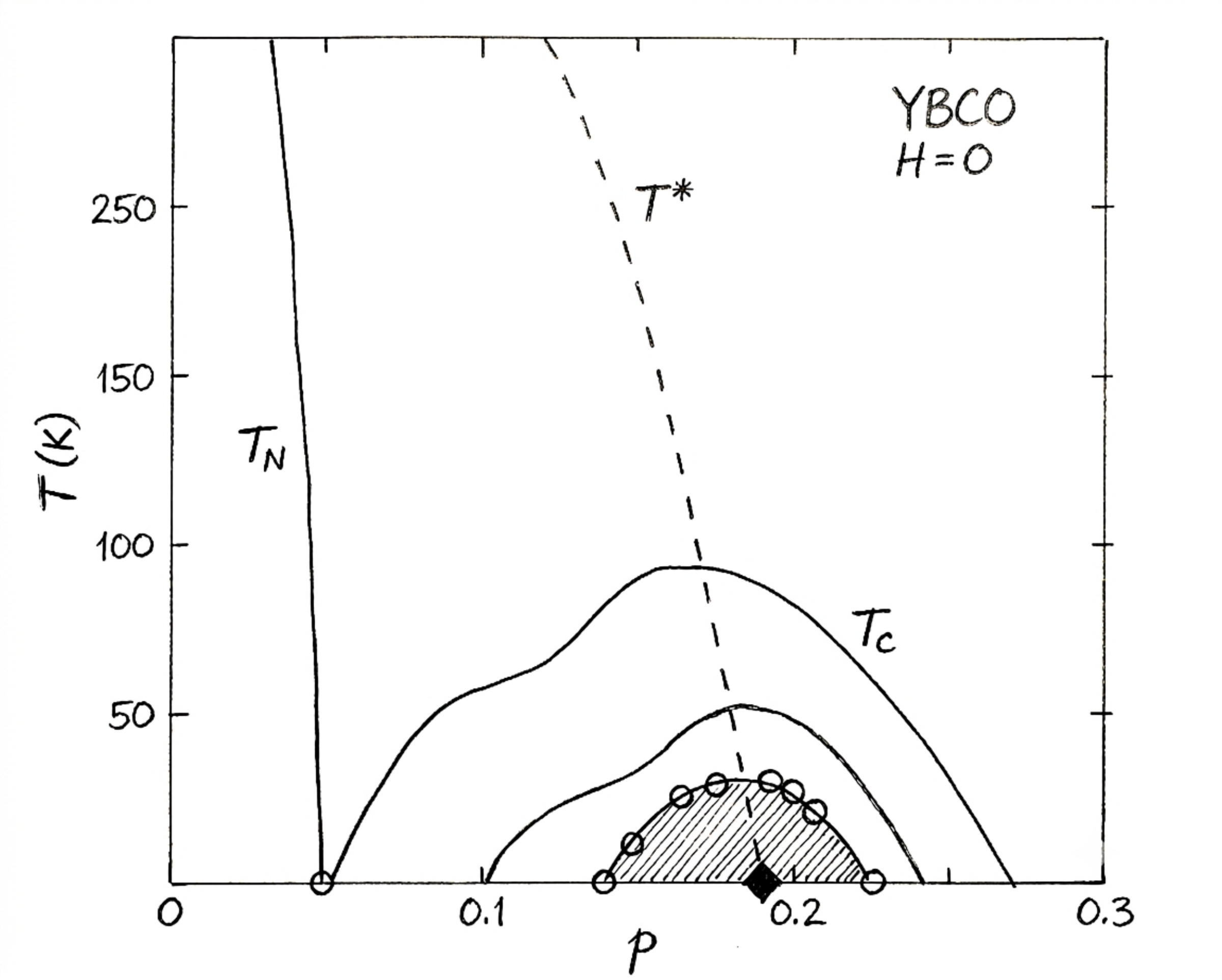}
    \caption{Schematic plot showing that the superconducting instability is the strongest near $p^*$! Reduced dome is due to $Zn$ doping (refer to\cite{bado} for details).}
  \label{fig4}
\end{figure}

\item As shown in figures (\ref{fig4}) the pairing tendencies are the strongest around $p^*$. As superconductivity is suppressed (partly) either by Zn doping or by strong magnetic field, the superconducting dome ``shrinks" around $p^*$. The present formulations of the mechanism either from repulsive scenario or from super-exchange scenario cannot address this pivotal experimental signature in a quantitative way.

\end{enumerate}

The rationale behind the newly added ``dogmas" (from (5) to (11)) is the following. The recognition that the ground state is a Fermi liquid (``dogma" 5) whereas the normal state at a finite temperature  (``dogma" 4) is a non-Fermi liquid clarifies the theoretical problem of connecting these two different ``fixed points". The models which assume a-priori non-Fermi liquid ground state are, therefore, ruled out. In the extended list of the ``dogmas", the author has added the Gor'kov-Teitel'baum and Barzykin-Pines phenomenologies, as these have been highly successful in rationalizing a wide variety of experimental facts (``dogmas" (8) and (9)). The ultimate  microscopic picture of the PG phase and that of the strange metal phase must respect these phenomenologies.

\renewcommand{\labelenumi}{\large \textcircled{\small\arabic{enumi}}.}

\section{Summary points}

\begin{enumerate}

\item The phenomenology of Barzykin and Pines gives us a comprehensive understanding of the spin degrees-of-freedom in the normal state. Within the PG phase the idea of two components (Spin liquid and a Fermi liquid) rationalizes nicely the magnetic susceptibility behaviour, and Johnston-Nakano scaling etc.  

\item The phenomenology of Gor'kov and Teitlel'baum gives us a comprehensive understanding as far as the charge degrees-of-freedom are concerned. The GTTA thermal activation gap in fact matches with the PG boundary. The thermal activation term itself goes to zero around $p^*\simeq 0.19$ which corroborates the observation of $p$ to $1+p$ transition, and destruction of the PG at $p^*$. 

\item The current formulations of the idea of superconductivity from repulsion and that of the super-exchange mechanism are not capable enough to address Taillefer's observation $A\propto T_c$.  And the answer to the question why superconducting instability around $p^*\simeq19$ is so strong is difficult to get from the present formulation of the super-exchange idea. 

\item In this author's opinion the theoretical deduction of  $A\propto T_c$,  and an answer to the question why superconducting instability around $p^*\simeq19$ is so strong constitutes a ``theoretical minimum". Once these are met, the problem of superconductivity in cuprates can be considered solved.

\end{enumerate}

In closing, the author would like to note that, even after four decades, cuprates continue to challenge, inspire, and engage the community. Their unresolved mysteries promise to keep the scientific community exploring the frontiers of correlated electron physics for many years to come. The journey is far from over, and the path remains illuminated by the incredible legacy of the past 40 years.

%
%
%
%
%
%


\acknowledgements{Statement regarding the use of AI: The plan, views, comments, and all the discussion presented above are by the author himself, ChatGPT is used to correct grammar and sentence construction at some places to improve the presentation, and to draw the sketches.}

\end{document}